\documentclass[12pt]{article}
\usepackage{amssymb}
\usepackage{epsfig}
\voffset= - 0.7in
\hoffset= - 0.6in         

\catcode`\@=11
\def\marginnote#1{}

\newcount\hour
\newcount\minute
\newtoks\amorpm
\hour=\time\divide\hour by60
\minute=\time{\multiply\hour by60 \global\advance\minute by-\hour}
\edef\standardtime{{\ifnum\hour<12 \global\amorpm={am}%
        \else\global\amorpm={pm}\advance\hour by-12 \fi
        \ifnum\hour=0 \hour=12 \fi
        \number\hour:\ifnum\minute<10 0\fi\number\minute\the\amorpm}}
\edef\militarytime{\number\hour:\ifnum\minute<10 0\fi\number\minute}

\def\draftlabel#1{{\@bsphack\if@filesw {\let\thepage\relax
   \xdef\@gtempa{\write\@auxout{\string
      \newlabel{#1}{{\@currentlabel}{\thepage}}}}}\@gtempa
   \if@nobreak \ifvmode\nobreak\fi\fi\fi\@esphack}
        \gdef\@eqnlabel{#1}}
\def\@eqnlabel{}
\def\@vacuum{}
\def\draftmarginnote#1{\marginpar{\raggedright\scriptsize\tt#1}}

\def\draft{\oddsidemargin -.5truein
        \def\@oddfoot{\sl preliminary draft \hfil
        \rm\thepage\hfil\sl\today\quad\militarytime}
        \let\@evenfoot\@oddfoot \overfullrule 3pt
        \let\label=\draftlabel
        \let\marginnote=\draftmarginnote
   \def\@eqnnum{(\theequation)\rlap{\kern\marginparsep\tt\@eqnlabel}%
\global\let\@eqnlabel\@vacuum}  }

\def\appname{Appendix}
\newcounter{app}
\def\theapp{\Alph{app}}
\def\app{\par
   \addvspace{4ex}
   \@afterindentfalse
  \secdef\@app\@dapp}
\def\@app[#1]#2{\ifnum \c@secnumdepth >\m@ne
        \refstepcounter{app}
        \addcontentsline{toc}{app}{\theapp
        \hspace{1em}#1}\else
      \addcontentsline{toc}{app}{ #1}\fi
   {\parindent \z@ \raggedright
    \Large \bf \appname~\theapp .
   \Large  \bf 
    #2}\nobreak
   \vskip 4ex   \noindent
\setcounter{equation}{0}
\def\theequation{\Alph{app}.\arabic{equation}}}
\def\@dapp#1{%
{\parindent \z@ \raggedright  \bf #1}\par\nobreak}
\def\l@app#1#2{\addpenalty{\@secpenalty}%
   \addvspace{1em plus\p@}%
   \begingroup
   \@tempdima 3em
     \parindent \z@ \rightskip \@pnumwidth
     \parfillskip -\@pnumwidth
     { \bf
     \leavevmode
     #1\hfil \hbox to\@pnumwidth{\hss #2}}\par
     \nobreak
   \endgroup}

\makeatletter
\newdimen\normalarrayskip            
\newdimen\minarrayskip               
\normalarrayskip\baselineskip
\minarrayskip\jot
\newif\ifold             \oldtrue            \def\new{\oldfalse}
\def\arraymode{\ifold\relax\else\displaystyle\fi}
\def\eqnumphantom{\phantom{(\theequation)}} 
\def\@arrayskip{\ifold\baselineskip\z@\lineskip\z@
     \else
     \baselineskip\minarrayskip\lineskip1\baselineskip\fi}

\def\@arrayclassz{\ifcase \@lastchclass \@acolampacol \or
\@ampacol \or \or \or \@addamp \or
   \@acolampacol \or \@firstampfalse \@acol \fi
\edef\@preamble{\@preamble
  \ifcase \@chnum
     \hfil$\relax\arraymode\@sharp$\hfil
     \or $\relax\arraymode\@sharp$\hfil
     \or \hfil$\relax\arraymode\@sharp$\fi}}

\def\@array[#1]#2{\setbox\@arstrutbox=\hbox{\vrule
     height\arraystretch \ht\strutbox
     depth\arraystretch \dp\strutbox
width\z@}\@mkpream{#2}\edef\@preamble{\halign \noexpand\@halignto
\bgroup \tabskip\z@ \@arstrut \@preamble \tabskip\z@ \cr}%
\let\@startpbox\@@startpbox \let\@endpbox\@@endpbox
  \if #1t\vtop \else \if#1b\vbox \else \vcenter \fi\fi
  \bgroup \let\par\relax
  \let\@sharp##\let\protect\relax
  \@arrayskip\@preamble}
\def\eqnarray{\stepcounter{equation}%
              \let\@currentlabel=\theequation
              \global\@eqnswtrue
              \global\@eqcnt\z@
              \tabskip\@centering              
              \let\\=\@eqncr
              $$%
            \halign to \displaywidth  \bgroup
             \eqnumphantom \@eqnsel
      \hskip\@centering                               
    $\displaystyle  \tabskip\z@ {##}$%
    &\global\@eqcnt\@ne \hskip 2\arraycolsep
         $ \displaystyle  \arraymode{##}$\hfil
    &\global\@eqcnt\tw@ \hskip 2\arraycolsep
         $\displaystyle\tabskip\z@{##}$\hfil
         \tabskip\@centering
    &{##}\tabskip\z@\cr}
\makeatother

\newfont{\hr}{msbm10}
\newfont{\ams}{msam10}

\font\numbers=cmss12
\font\upright=cmu10 scaled\magstep1
\def\stroke{\vrule height8pt width0.4pt depth-0.1pt}
\def\topfleck{\vrule height8pt width0.5pt depth-5.9pt}
\def\botfleck{\vrule height2pt width0.5pt depth0.1pt}
\def\Zmath{\vcenter{\hbox{\numbers\rlap{\rlap{Z}\kern 0.8pt\topfleck}\kern
2.2pt
                   \rlap Z\kern 6pt\botfleck\kern 1pt}}}
\def\Qmath{\vcenter{\hbox{\upright\rlap{\rlap{Q}\kern
                   3.8pt\stroke}\phantom{Q}}}}
\def\Nmath{\vcenter{\hbox{\upright\rlap{I}\kern 1.7pt N}}}
\def\Cmath{\vcenter{\hbox{\upright\rlap{\rlap{C}\kern
                   3.8pt\stroke}\phantom{C}}}}
\def\Rmath{\vcenter{\hbox{\upright\rlap{I}\kern 1.7pt R}}}
\def\Z{\ifmmode\Zmath\else$\Zmath$\fi}
\def\Q{\ifmmode\Qmath\else$\Qmath$\fi}
\def\N{\ifmmode\Nmath\else$\Nmath$\fi}
\def\C{\ifmmode\Cmath\else$\Cmath$\fi}
\def\R{\ifmmode\Rmath\else$\Rmath$\fi}

\def\d{\partial}

\def\bea{\begin{eqnarray}}
\def\eea{\end{eqnarray}}

\def\beq{\begin{equation}}
\def\eeq{\end{equation}}
\def\ba{\beq\new\begin{array}{c}}
\def\ea{\end{array}\eeq}
\def\be{\ba}
\def\ee{\ea}

\def\e{\epsilon}

\def\half{{\textstyle{1\over2}}}

\def\N2{${\cal N}=2$}
\def\4N{${\cal N}=4$}
\def\1N{${\cal N}=1$}
\def\1N*{${\cal N}=1^*$}

\def\beq{\begin{equation}}
\def\eeq{\end{equation}}
\def\ba{\beq\new\begin{array}{c}}
\def\ea{\end{array}\eeq}
\def\be{\ba}
\def\ee{\ea}
\newcommand{\rf}[1]{(\ref{#1})}

\begin{document}



\setcounter{footnote}{3}
\renewcommand{\thefootnote}{\fnsymbol{footnote}}
\begin{center}
{\Large\bf
First Order String Theory and\\
\vspace{0.3 cm}
the Kodaira-Spencer Equations. I
}\\
\vspace{1.5 cm}
{\large O.~Gamayun\footnote{
Bogolyubov Institute of Theoretical Physics, Kiev, Ukraine},
A.~S.~Losev\footnote{Institute of Systems Research and
Institute for Theoretical and Experimental Physics, Moscow, Russia},
A.~Marshakov\footnote{
Lebedev Physics Institute and
Institute for Theoretical and Experimental Physics, Moscow, Russia}}\\
\vspace{0.6 cm}
\end{center}
\vspace{0.3 cm}
\begin{quotation}
\noindent
We consider first-order bosonic string theory, perturbed by the primary
operator, corresponding to deformation of the target-space complex structure.
We compute the effective action in this theory and find that its consistency
with the world-sheet conformal invariance requires necessarily the
Kodaira-Spencer equations to be satisfied by target-space Beltrami differentials.
We discuss the symmetries of the theory and its reformulation
in terms of the vielbein background fields.
\end{quotation}
\renewcommand{\thefootnote}{\arabic{footnote}}
\setcounter{section}{0} \setcounter{footnote}{0}
\setcounter{equation}0

\section{Introduction}

The formulation of string theory in non-trivial background is a long-standing nontrivial
problem (see e.g. \cite{Pol,GSW}). The most traditional approach \cite{FT,CFoth} is based
almost totally on studying the two-dimensional sigma-models, where many issues
directly concerning the basic principles of string theory are hidden behind the serious
technical problems of computations in nonlinear interacting theories. Even the heart
of perturbative string theory -
the concept of two-dimensional conformal invariance - is not directly seen within this approach,
since the theory is conformally-invariant only on mass shell, which basically forbids to consider
the most part of the interesting deformations of the target-space background.

In order to try to avoid at least some of these complications, it has been proposed in \cite{LMZ}
to study the simplest possible string model - the first-order string theory with the ``bare action''
of the bosonic first-order free conformal field theory
\be
\label{act}
S_0=\frac{1}{2\pi\alpha'}\int_\Sigma d^2 z (p_i\bar{\partial} X^{i}+
p_{\bar{i}}{\partial} X^{\bar{i}})
\ee
which is independent of the target-space metric and requires only some (local) choice
of the complex structure. Analogy with the Hamiltonian formalism in the theory of particle
suggests that \rf{act} corresponds at least naively to a background with the singular
target-space metric (and the singular Kalb-Ramond $B$-field \cite{LMZ}). The world-sheet
fields $\{ X^\mu \} = \{\left(X^i,X^{\bar i}\right)\}$,
$\{ p_i \}$ and $\{ p_{\bar i} \}$ (with $\mu=1,\ldots,D$; $i,{\bar i}=1,\ldots,D/2$) are
sections of $H^0(\Sigma)$, $H^{(1,0)}(\Sigma)$ and $H^{(0,1)}(\Sigma)$ correspondingly,
being holomorphic
(or anti-holomorphic) on the equations of motion. The only nontrivial operator product
expansions (OPE) for the theory \rf{act} are
\be
\label{px}
p_i(z) X^j(z') = \frac{\alpha'\delta_i^j}{z-z'} + {\rm regular\ terms}
\ee
together with their complex conjugated.

The free field theory action \rf{act} can be naturally perturbed by the operators
\be
\label{vg}
V_g = \frac{1}{2\pi\alpha'}\int_\Sigma O_g =
\frac{1}{2\pi\alpha'}\int_\Sigma d^2z g^{i\bar{j}}p_i p_{\bar{j}}
\ee
with the $X$-dependent ``coefficient functions'' or target-space fields $g^{i\bar{j}} =
g^{i\bar{j}}(X)$, as well as
\be
\label{vmu}
V_\mu = \frac{1}{2\pi\alpha'}\int_\Sigma O_\mu =
\frac{1}{2\pi\alpha'}\int_\Sigma d^2z \mu_{\bar{i}}^j\bar{\partial} X^{\bar{i}} p_j
\\
V_{\bar\mu} = \frac{1}{2\pi\alpha'}\int_\Sigma O_{\bar\mu} =
\frac{1}{2\pi\alpha'}\int_\Sigma d^2z \bar{\mu}^{\bar{j}}_i\partial X^i p_{\bar{j}}
\ee
where $\mu_{\bar{i}}^j = \mu_{\bar{i}}^j(X)$ (together with its complex conjugated
${\bar\mu}^{\bar{i}}_j = {\bar\mu}^{\bar{i}}_j(X)$), and
\be
\label{vb}
V_b = \frac{1}{2\pi\alpha'}\int_\Sigma O_b =
\frac{1}{2\pi\alpha'}\int_\Sigma d^2z b_{i\bar{j}}\partial X^i\bar{\partial} X^{\bar{j}}
\ee
where again $b_{i\bar{j}} = b_{i\bar{j}}(X)$. It is also often useful to define
the ``real'' operator
\be
\label{Fimu}
\Phi(z,{\bar z}) = O_\mu(z,{\bar z}) + O_{\bar\mu}(z,{\bar z}) =
\mu_{\bar{i}}^j\bar{\partial} X^{\bar{i}} p_j
+ \bar{\mu}^{\bar{j}}_i\partial X^i p_{\bar{j}}
\ee
In order for the operators \rf{vg},\rf{vmu} and \rf{Fimu} to be well-defined as
conformal primary operators, one has to impose the transversality conditions for
the background fields
\be
\label{transv}
\d_i g^{i{\bar j}} = 0,\ \ \ \d_{\bar j} g^{i{\bar j}} =0
\\
\d_i \mu_{\bar{j}}^i = 0,\ \ \ \d_{\bar j} \bar{\mu}^{\bar{j}}_i  =0
\ee
which allow to get rid of the singularities, possibly arising from ``internal'' contractions in
\rf{vg},\rf{vmu} and \rf{Fimu} or, in different words, the higher-order poles
in the operator-product expansions with the components of the
stress-energy tensor $T \sim p_i\d X^i$ and ${\bar T} \sim p_{\bar i}{\bar\d} X^{\bar i}$ in the
bare theory \rf{act}.

The operators \rf{vg}-\rf{vb} (or \rf{Fimu}) are the only possible
marginal $(\Delta,{\bar\Delta})=(1,1)$
primary operators in the first-order theory \rf{act}. In addition, one can also introduce the
holomorphic $(1,0)$-currents
\be
\label{cur}
j_v = p_iv^i(X),\ \ \ \d_iv^i=0
\\
j_\omega = \omega_i(X)\d X^i
\ee
(and their anti-holomorphic $(0,1)$-conjugates),
which generate the holomorphic change of co-ordinates and gauge transformations (their
anomalous operator algebra has been studied in \cite{MSV,W,LMZ,N}). The non-holomorphic
operators, similar to \rf{cur}, can also arise when studying generic non-holomorphic
symmetries of the perturbed action, and these symmetries will be partly considered
below.

We
are going to study the conditions, when the operators \rf{vg}-\rf{vb} become exactly
marginal or can be raised up to the exponent and added to the free action \rf{act}.
In other words, this is equivalent to vanishing of their beta-functions
in the perturbed theory \cite{Pol,Zam}.
The quadratic (in background fields) contributions to these beta-functions are
given by the structure constants of the OPE's of the primary operators \rf{vg}-\rf{vb},
whose vanishing leads, for example, to the nonlinear equation
\be
\label{geq}
g^{i\bar{j}}\d_i\d_{\bar j}g^{k\bar{l}}-\d_ig^{k\bar{j}}\d_{\bar j}g^{i\bar{l}}=0
\ee
for the functions $g^{i\bar{j}}(X)$, shown to
be a direct analog of the Einstein equations for the physical fields $G$, $B$ and $\Phi$ (the
target-space metric, the Kalb-Ramond antisymmetric two-form and the dilaton), related
with the background fields $g$, $\mu$, ${\bar\mu}$ and $b$ from \rf{vg}, \rf{vmu} and \rf{vb}
by a nontrivial transformation \cite{LMZ}.
In the present paper we would like to concentrate mostly on the background equations of motion
for the target-space
``Beltrami'' fields\footnote{
These fields in the context of Lagrangian field theory were discussed already in
\cite{Lazar}, the Beltrami parametrization of the world-sheet geometry in string theory
was discussed e.g. in \cite{Baul}.} $\mu = dX^{\bar{j}}\mu_{\bar{j}}^i{\d \over\d X^i}$
and $\bar\mu = dX^i \bar{\mu}^{\bar{j}}_i{\d \over\d X^{\bar{j}}}$, keeping the other fields
to be shut down for a while, or playing maximally a role of a ``spectator'' or
``probe'' operators.
In such case the vertex operators \rf{vmu} and \rf{Fimu} can be obviously considered as
deforming the complex structure of the original bare theory \rf{act}, and from generic
target-space symmetry reasons one would expect that the corresponding fields should
satisfy the Kodaira-Spencer equations \cite{KoSpe}
\be
\label{KoSpe}
N_{{\bar k}{\bar j}}^i \equiv \d_{[{\bar k}}\mu^i_{{\bar j}]} -
\mu_{[\bar k}^l\d_l\mu_{\bar j]}^i =0
\\
{\bar N}_{ik}^{\bar j} \equiv \d_{[i}{\bar\mu}^{\bar j}_{k]} -
{\bar\mu}_{[i}^{\bar l}\d_{\bar l}{\bar\mu}_{k]}^{\bar j} = 0
\ee
which have an obvious sense of vanishing of the Nijenhuis tensor or
curvatures for the gauge fields
$\mu = dX^{\bar{j}}\mu_{\bar{j}}^i{\d \over\d X^i}$ and
$\bar\mu = dX^i \bar{\mu}^{\bar{j}}_i{\d \over\d X^{\bar{j}}}$ with the values in Lie algebra
of the vector fields in tangent bundle to the target manifold (see \cite{BCOV} for brief
description of Kodaira-Spencer theory and their important applications for topological
strings). Below we are going to derive these equations directly from
the consistency of perturbed first-order conformal field theory.

\section{Background field expansion
\label{ss:bfe}}

The most common analysis of the beta-functions follows from studying logarithmic divergences
in the effective action, coming from its one-loop computation (see various issues of
this procedure e.g. in \cite{FT,GSW,ZJ}).
Consider the first-order theory with the action $S = S_0 + \int_{\Sigma}\Phi$ and
decompose the would-sheet fields into the fast and slow (or quantum and classical) parts
$X \to X_{\rm cl}+ \sqrt{\alpha'}X$ and $p \to p^{\rm cl} + \sqrt{\alpha'}p$.

Expanding the Lagrangian up to the second order one gets
\be
\label{Lexp}
{\cal L} = {\cal L}_0 + \Phi  = {\cal L}_{\rm cl} + \alpha'\left( p_i {\bar \d}{\tilde X}^i +
p_i{\bar \d}X_{\rm cl}^{\bar j}\d_k\mu^i_{\bar j}(X_{\rm cl}) {\tilde X}^k +
p_i{\bar \d}X_{\rm cl}^{\bar j}N_{{\bar k}{\bar j}}^i(X_{\rm cl}) X^{\bar k} +
c.c. \right) +o(\alpha')
\ee
where
\be
\label{tX}
{\tilde X}^i = X^i + \mu^i_{\bar k}(X_{\rm cl})X^{\bar k}
\ee
and the Kodaira-Spencer term $N_{{\bar k}{\bar j}}^i$ has been already defined
in \rf{KoSpe}. The change of quantum co-ordinates \rf{tX} is an obvious
transformation of the target-space complex polarization
$\{ X^{i},X^{\bar i}\}\rightarrow \{ {\tilde X}^i,{\tilde X}^{\bar i}\}$,
caused by expansion around the classical background $X_{\rm cl}$ instead of
the zero background, originally taken if directly dealing with the bare action \rf{act}.

The linear in fluctuations terms disappear from \rf{Lexp} due
to the equations of motion
\be
\label{eqmox}
{\bar\d} X_{\rm cl}^i + \mu^i_{\bar j}(X_{\rm cl}){\bar\d} X_{\rm cl}^{\bar j} =0
\ee
together with its complex conjugated, and
\be
\label{eqmop}
\left({\bar\nabla} p^{\rm cl}\right)_k +
{\bar\mu}^{\bar i}_k(X_{\rm cl})\left(\nabla p^{\rm cl}\right)_{\bar i} =
p^{\rm cl}_{\bar i}\d X_{\rm cl}^j{\bar N}^{\bar i}_{kj}(X_{\rm cl})
\ee
with
\be
\label{nap}
\left({\bar\nabla} p^{\rm cl}\right)_k =
{\bar\d}p^{\rm cl}_k - \d_k\mu^i_{\bar j}(X_{\rm cl}){\bar\d} X_{\rm cl}^{\bar j}p^{\rm cl}_i
\ee
Note, that the combination \rf{KoSpe} arises already in the r.h.s. of the
{\em classical} equations for momenta, which
turn into $\left({\bar\nabla} p^{\rm cl}\right)_k =0$ together with its complex
conjugated for the background fields $\mu(X_{\rm cl})$ obeying the Kodaira-Spencer
equations.

After the change of variables, inverse to \rf{tX}, one gets for the ``old co-ordinates''
\be
\label{xxt}
X^{\bar k} = \left((1-{\bar\mu}\mu(X_{\rm cl}))^{-1}\right)^{\bar k}_{\bar l}{\tilde X}^{\bar l}
- {\bar\mu}^{\bar k}_s(X_{\rm cl})\left((1-\mu{\bar\mu}(X_{\rm cl}))^{-1}\right)^s_l {\tilde X}^l
=
\\
= {\bar M}^{\bar k}_{\bar l}(X_{\rm cl}){\tilde X}^{\bar l} - {\bar\mu}^{\bar k}_s(X_{\rm cl})
M^s_l(X_{\rm cl}) {\tilde X}^l
\ee
where
\be
\label{Mmatr}
M^i_j = \left(\delta^i_j-(\mu\bar\mu)^i_j\right)^{-1}
\ee
(together with the corresponding complex conjugated formulas), and the
Lagrangian \rf{Lexp} acquires the form
\be
\label{Lexpxt}
{\cal L} = {\cal L}_0 + \Phi  = {\cal L}_{\rm cl} + \alpha'\left( p_i {\bar \d}{\tilde X}^i +
p_{i}U^{i}_{\bar{j}}{\tilde X}^{\bar{j}}+
p_{i}W^{i}_{j}{\tilde X}^{j} + c.c\right) + o(\alpha')
\ee
where
\be
\label{vertV}
U^{i}_{\bar{j}} = N_{{\bar k}{\bar l}}^i
{\bar M}^{\bar k}_{\bar j}(X_{\rm cl}){\bar \d}X_{\rm cl}^{\bar l}
\ee
and
\be
\label{vertW}
W^{i}_{j}={\bar \d}X_{\rm cl}^{\bar j}\left(\d_j\mu^i_{\bar j}(X_{\rm cl})
- N_{{\bar k}{\bar j}}^i{\bar\mu}^{\bar k}_s
M^s_j(X_{\rm cl}) \right)
\ee
(together with their complex conjugated)
are two dependent on external fields $X_{\rm cl}$ set of vertices, to be treated as
perturbation of the free field theory
${\tilde{\cal L}}_0 = p_i {\bar \d}{\tilde X}^i + c.c.$ with the propagators
\begin{figure}[tp]
\epsfysize=2.5cm
\centerline{\epsfbox{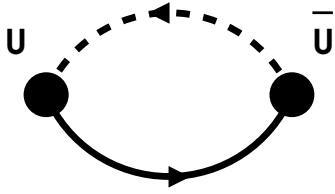}}
\caption{The only 2-vertex 1-loop diagram which diverges
logarithmically. Black and dash lines denote the holomorphic and anti-holomorphic propagators
from \rf{ptxprop}.}
\label{fi:2mom}
\end{figure}
\be
\label{ptxprop}
\langle p_i(z){\tilde X}^j(z')\rangle = {\delta_i^j\over z-z'} =
\delta_i^j\int {d^2q\over (2\pi)^2}{e^{i{\bf q}({\bf z}-{\bf z}')}\over {\bar q}}
\\
\langle p_{\bar i}(z){\tilde X}^{\bar j}(z')\rangle = {\delta_{\bar i}^{\bar j}
\over {\bar z}-{\bar z}'} =
\delta_{\bar i}^{\bar j}\int {d^2q\over (2\pi)^2}{e^{i{\bf q}({\bf z}-{\bf z}')}\over q}
\ee
An important observation is that both vertices \rf{vertV} and \rf{vertW}
do {\em not} depend on the momenta of quantum fields $X$ and $p$. Therefore,
the only logarithmically divergent contribution to the effective action comes from the
paring of $U^{i}_{\bar{j}}$ and $\bar{U}_{i}^{\bar{j}}$ vertices or from the
diagram depicted at fig.~\ref{fi:2mom}, which gives the contribution
\be
\label{divdi}
\Gamma^{\rm div} \sim \int d^2z\ \frac{U^{i}_{\bar{j}}(z)\bar{U}_{i}^{\bar{j}}(0)}{|z|^2}
\sim U^{i}_{\bar{j}}\bar{U}_{i}^{\bar{j}}\int {d^2q\over q{\bar q}}
\ee
The logarithmically divergent integrals in \rf{divdi} can be just cut off, say by $|z|> \epsilon$,
which leads to the renormalization of the operator of type \rf{vb}
\be
\delta b_{i\bar{j}} \sim \log\epsilon \cdot B_{i\bar{j}}
\ee
where
\be\label{betab}
B_{i{\bar j}} =
-N^l_{{\bar k}{\bar j}}{\bar N}^{\bar l}_{ki}M^k_l{\bar M}^{\bar k}_{\bar l}
\equiv B^{(2)}_{i{\bar j}} +B^{(3)}_{i{\bar j}}+\bar{B}^{(3)}_{i{\bar j}} + O(\mu^4) =
\\
=-\d_{[k}{\bar\mu}^{\bar k}_{i]}\d_{[\bar{k}} \mu^k_{\bar{j}]}
 + \d_{[k}{\bar\mu}^{\bar k}_{i]}\mu^l_{[\bar{k}}\d_l\mu^i_{\bar{j}]}
+ \d_{[\bar{k}} \mu^k_{\bar{j}]}\bar{\mu}^{\bar{l}}_{[k}\d_{\bar{l}}\bar{\mu}^{\bar{k}}_{i]}+
O(\mu^4)
\ee
obviously vanishing on the solutions to \rf{KoSpe}.
The Jacobian of transformation \rf{tX} which changes the measure in the path integral is
derivative independent, and therefore does not affect the result of computation of the one-loop
diagram in the theory \rf{Lexpxt}. Actually, this result is even exact in all orders in $\alpha'$,
since any vertex, obtained by the $\alpha'$ expansion \rf{Lexp}, has a single $p$ ($\bar{p}$)
together with many $\tilde{X}$ ($\tilde{\bar{X}}$) legs, so that one cannot in principle construct
a diagram with more than one loops.

\begin{figure}[tp]
\epsfysize=2.5cm
\centerline{\epsfbox{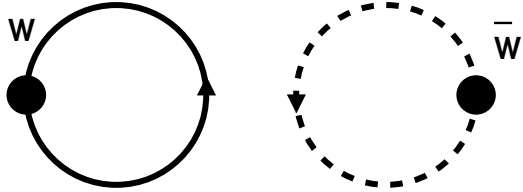}}
\caption{The 1-loop tadpole diagrams which give rise naively to a linear divergences.
Black and dash lines again denote the holomorphic and anti-holomorphic propagators
from \rf{ptxprop}.}
\label{fi:tadpole}
\end{figure}

Finally in this section let us point out, that vertices \rf{vertW} could be perhaps neglected,
if the current $j_v$ from \rf{cur} is not anomalous: they can produce only the
linearly divergent
tadpole diagrams (see fig.~\ref{fi:tadpole}), where the divergency is killed by the angle integration.
However, since ${\bar\d}\langle j_v\rangle =
{1\over 2\pi}R^{(2)}\d_iv^i$, the presence of vertices \rf{vertW} in the Lagrangian leads to creation
of the terms
\be
\label{anoct}
W^{i}_{j}\langle p_{i}{\tilde X}^{j}\rangle \sim
{\bar \d}X_{\rm cl}^{\bar j} N_{{\bar k}{\bar j}}^i{\bar\mu}^{\bar k}_s
M^s_j(X_{\rm cl}) = {\bar \d}X_{\rm cl}^{\bar j}
N_{{\bar k}{\bar j}}^i{\bar\mu}^{\bar k}_i(X_{\rm cl}) + O(\mu^4)
\ee
We shall discuss in the second part of this paper \cite{GM2}, that extra terms
like \rf{anoct} are related to the higher
singularities in the operator product expansions and can give rise to extra singularities
in the correlation functions.

\section{Symmetries}

The bare action \rf{act} is invariant under the holomorphic change of variables
\be
\label{holsym}
X^{i} \rightarrow X^{i} - v^{i}(X)
\\
p_{i} \rightarrow p_{i} + \d_iv^k(X)p_k
\ee
generated by the first current $j_v$ from \rf{cur}, together with their complex conjugated.
In order to study the general co-ordinate invariance of the theory one needs to switch on all
background fields \rf{vg}, \rf{vmu} or \rf{Fimu}, and \rf{vb}. The general co-ordinate transformations
of the fundamental variables of the first-order theory should be then supplemented by the transformations
of the background fields.

The perturbed action
\be
\label{pertact}
S = S_0 + {1\over 2\pi\alpha'}\int d^2z\left(O_g + O_\mu + O_{\bar\mu} + O_b\right) =
\\
= {1\over 2\pi\alpha'}\int d^2z \left(p_i\bar{\d} X^{i} +
p_{i}\mu^{i}_{\bar{j}}\bar{\d} X^{\bar{j}} + c.c.\right) +
{1\over 2\pi\alpha'}\int d^2z \left(g^{i\bar{j}}p_ip_{\bar{j}} +
b_{i{\bar j}}{\d} X^{i}\bar{\d} X^{\bar{j}}\right)
\ee
is invariant already under the transformation of co-ordinates
\be
\label{gencoord}
X^{i} \rightarrow X^{i} - v^{i}(X,\overline{X}) ,\,\,\,\,\,\,\,\,
X^{\overline{j}} \rightarrow X^{\overline{j}} - v^{\overline{j}}(X,\overline{X})
\ee
where the momenta transform already in the background-field dependent way
\be
p_{i} \rightarrow p_{i} + p_k(\d_iv^k + \mu^k_{\bar{k}}\d_iv^{\bar{k}})
+ b_{j{\bar k}}\d_i v^{\bar k}\d X^j
\ee
together with the complex-conjugated formula.
For the background fields themselves one gets
\be\label{mutr}
\mu^{i}_{\bar{j}} \rightarrow \mu^{i}_{\bar{j}} +
\d_{\bar{j}}v^i + v^{k}\d_k\mu^{i}_{\bar{j}} +
v^{\bar{k}}\d_{\bar{k}}\mu^{i}_{\bar{j}}
+\mu^{i}_{\bar{k}}\d_{\bar{j}}v^{\bar{k}} -
\mu^k_{\bar{j}}\d_kv^i
 - \mu^i_{\bar{k}}\mu^k_{\bar{j}}\d_k v^{\bar{k}} - g^{i{\bar k}}b_{l{\bar j}}\d_{\bar k}v^l=
\\
=\mu^{i}_{\bar{j}} + \d_{\bar{j}}v^i + \{ v, \mu \}^i_{\bar j} +
v^{\bar{k}}\d_{\bar{k}}\mu^{i}_{\bar{j}} +\mu^{i}_{\bar{k}}\left(\d_{\bar{j}}
- \mu^k_{\bar{j}}\d_k\right) v^{\bar{k}} - g^{i{\bar k}}b_{l{\bar j}}\d_{\bar k}v^l
\ee
where
\be
\label{shou}
\{ v, \mu \}^i_{\bar j} = v^{k}\d_k\mu^{i}_{\bar{j}} - \mu^k_{\bar{j}}\d_kv^i
\ee
is the Schouten bracket, or the analog of the commutator for the gauge connections with values
in the Lie algebra of vector fields, and also
\be
\label{gtr}
g^{i{\bar j}} \rightarrow g^{i{\bar j}} + v^k\d_k g^{i{\bar j}} + v^{\bar{k}}\d_{\bar{k}} g^{i{\bar j}}
- g^{i{\bar k}}(\d_{\bar k}v^{\bar j}+{\bar\mu}^{\bar j}_k\d_{\bar k}v^k)
- g^{k{\bar j}}(\d_kv^i+\mu^{i}_{\bar{k}}\d_{k}v^{\bar{k}})
\ee
and
\be
\label{btr}
b_{i{\bar j}} \rightarrow b_{i{\bar j}} + v^k\d_k b_{i{\bar j}} + v^{\bar{k}}\d_{\bar{k}} b_{i{\bar j}}
+ b_{i{\bar k}}(\d_{\bar{j}}-\mu^{k}_{\bar j}\d_k)v^{\bar{k}}
+ b_{l{\bar j}}(\d_i-{\bar\mu}^{\bar k}_i\d_{\bar k})v^l
\ee
These transformations almost coincide with conventional Lie derivatives - they partially
differ from naively expected target-space transformation law by replacing
\be
\label{vrepl}
\d_kv^i \rightarrow \d_kv^i+\mu^{i}_{\bar{k}}\d_{k}v^{\bar{k}}
\\
\d_{\bar{j}}v^{\bar{k}} \rightarrow (\d_{\bar{j}}-\mu^{k}_{\bar j}\d_k)v^{\bar{k}} =
\mathcal{D}_{\bar j}v^{\bar{k}}
\ee
or introducing, in the second line, the ``long derivatives''
\be
\label{longder}
\d_{\bar i} \to \mathcal{D}_{\bar i}=\d_{\bar i} -\mu^k_{\bar i}\d_k,
\ \ \ \
[\mathcal{D}_{\bar i},\mathcal{D}_{\bar j}] = -N^k_{\bar i\bar j}\d_k
\ee
The sense of this replacement can be understood as follows: consider for simplicity constant
background $\mu$-field, then when it is switched on the correct target-space holomorphic
co-ordinate becomes ${\hat X}^i=X^i+\mu^i_{\bar j}X^{\bar j}$ (cf. also with \rf{tX}).
Its infinitesimal variation ${\hat v}^i=v^i+\mu^i_{\bar j}v^{\bar j}$ under the
transformation \rf{gencoord} stays in the first line of \rf{vrepl}, while the corresponding
$\bar\d$-operator \rf{longder} (such that
$\mathcal{D}_{\bar j}{\hat X}^i=0$) appears in the second line of \rf{vrepl}.

Transformation \rf{mutr} (at vanishing $b$-field $b_{i{\bar j}}=0$)
induces the following transformation of the tensor
$N_{{\bar k}{\bar j}}^i \equiv \d_{[{\bar k}}\mu^i_{{\bar j}]} -
\mu_{[\bar k}^l\d_l\mu_{\bar j]}^i$
\be
\label{Ntrans}
N_{{\bar s}{\bar j}}^i \rightarrow N_{{\bar s}{\bar j}}^i - (\d_kv^i+\mu^{i}_{\bar{k}}\d_{k}v^{\bar{k}})
N_{{\bar s}{\bar j}}^k
+ (\d_{\bar{s}}v^{\bar{k}}-\mu^{k}_{\bar s}\d_kv^{\bar{k}})N_{{\bar k}{\bar j}}^i
 + (\d_{\bar{j}}v^{\bar{k}}-\mu^{k}_{\bar j}\d_kv^{\bar{k}})N_{{\bar s}{\bar k}}^i
\\
 +  v^{k}\d_{k}N_{{\bar s}{\bar j}}^i+ v^{\bar{k}}\d_{\bar{k}}N_{{\bar s}{\bar j}}^i
\ee
The transformation laws \rf{mutr}, \rf{Ntrans} generate the following covariant behavior of the
beta-function \rf{betab}
\be
\label{covbeta}
B_{i{\bar j}} \rightarrow B_{i{\bar j}} + v^k\d_k B_{i{\bar j}} + v^{\bar{k}}\d_{\bar{k}} B_{i{\bar j}}
+ B_{i{\bar k}}(\d_{\bar{j}}-\mu^{k}_{\bar j}\d_k)v^{\bar{k}}
+ B_{l{\bar j}}(\d_i-{\bar\mu}^{\bar k}_i\d_{\bar k})v^l
\ee
which, according to consistency requirements, exactly coincide with the transformation
law for the $b$-field \rf{btr}. It is necessary to point out, that formulas \rf{covbeta}
and \rf{btr} coincide exactly, only when the nontrivial transformation law of \rf{Mmatr}
is taken into account. Also, multiplying by the matrices $M^i_j$ and
$\bar{M}^{\bar{i}}_{\bar{j}}$ from \rf{Mmatr}, one can rewrite e.g.
\rf{gtr}, \rf{Ntrans}
\be
\label{gNtrans}
\widehat{g}^{i\bar{j}} \to \widehat{g}^{i\bar{j}} + v^k\d_k \widehat{g}^{i{\bar j}} +
v^{\bar{k}}\d_{\bar{k}} \widehat{g}^{i{\bar j}}
 - \mathcal{D}_{k}v^{i}\widehat{g}^{k\bar{j}} -
 \mathcal{D}_{\bar{k}}v^{\bar{j}}\widehat{g}^{i\bar{k}}
\\
\widehat{N}_{{\bar s}{\bar j}}^i \rightarrow \widehat{N}_{{\bar s}{\bar j}}^i -
\mathcal{D}_{k}v^i\widehat{N}_{{\bar s}{\bar j}}^k
+ \mathcal{D}_{{\bar s}}v^{\bar{k}}\widehat{N}_{{\bar k}{\bar j}}^i
 + \mathcal{D}_{{\bar j}}v^{\bar{k}}\widehat{N}_{{\bar j}{\bar k}}^i
 +  v^{k}\d_{k}\widehat{N}_{{\bar s}{\bar j}}^i+ v^{\bar{k}}\d_{\bar{k}}\widehat{N}_{{\bar s}{\bar j}}^i
\ee
for $\widehat{g}^{i\overline{j}} = M^i_kM^{\overline{j}}_{\overline{k}}g^{k\overline{k}}$ and
$\widehat{N}_{{\bar s}{\bar j}}^i = M^{i}_kN_{{\bar s}{\bar j}}^k$
purely in terms of the long derivatives \rf{longder}. Not quite canonical form of the field's
transformations found in this section can be explained by considering perturbed first-order
theory as a particular gauge of more generic form for such action, written in terms of
the vielbein background fields.

\section{The alternative first-order action
\label{ap:alt}}

The perturbed action \rf{Lexp} can be also considered as a particular case of more general
non-linear theory with the action
\be
\label{foa}
L= p_a e_{\mu}^{a}(X) \bar{\partial} X^\mu + c.c + \ldots
\ee
where $a,b = 1,\ldots,D/2$ and $\mu,\nu=  1,\ldots,D$, and
the external fields appear now in the form of veilbein, or
the one-forms $e = e_\mu dX^\mu$ and $\bar{e}=\bar{e}_\mu dX^\mu$ with values in the
holomorphic and antiholomorphic subspaces of the complexified tangent bundle. The covariance
of the action \rf{foa} is transparent, while the action \rf{Lexp} can be obtained from
\rf{foa} imposing the gauge $e^a_i = \delta^a_i$, $e^a_{\bar j}=\mu^a_{\bar j}$.
We also define ($A=\{a,{\bar a}\}$)
\be
\label{inviel}
e^\mu_A e^B_\mu = \delta_A^B,\ \ \ A,B=1,\ldots,D,\ \ \ {\rm or}
\\
e^\mu_a e^b_\mu = \delta_a^b,\ \ \ a,b=1,\ldots,D/2
\\
e^\mu_{\bar a} e^{\bar b}_\mu = \delta_{\bar a}^{\bar b},\ \ \ {\bar a},{\bar b}=1,\ldots,D/2
\ee
The fields $p_a$  and $p_{\bar{a}}$ are still
$(1,0)$ and  $(0,1)$ world-sheet forms with values in the
pullbacks of the holomorphic and antiholomorphic pieces of the cotangent
bundle, equipped with the unitary structure.

Decomposing again $X\rightarrow X+x$ into classical fields and quantum fluctuations,
one gets for \rf{foa}
\be
L= \ldots + p_a e_{\mu}^{a}(X) \bar{\partial} x^\mu +
p_a\d_\nu e_{\mu}^{a}(X)\d X^\mu x^\nu + c.c + \ldots
\ee
which can be further rewritten introducing $x^\mu = e^\mu_a(X)Y^a+e^\mu_{\bar a}(X)Y^{\bar a}$
as
\be
\label{foa2}
L = \ldots + p_a \bar{\partial} Y^a+  p_{a} {\cal W}_{b}^{a}Y^b +
 p_{a} {\cal U}_{\bar{b}}^{a}\bar{Y}^{\bar{b}} + c.c + \ldots
\ee
with
\be
\label{bae}
{\cal U}_{\bar{b}}^{a}=(de)_{\mu\nu}^{a} e_{\bar{b}}^{\mu}
\bar{\partial} X^{\nu},\ \ \
{\cal W}_{b}^{a}=(de)_{\mu\nu}^{a} e_{b}^{\mu} \bar{\partial} X^{\nu}
\\
(de)^a_{\mu\nu} = \d_{[\mu}e^a_{\nu]}
\ee
Following the same argumentation, as in sect.~\ref{ss:bfe}, one gets that the
logarithmic divergence comes from the only one-loop diagram (depicted at fig.~\ref{fi:2mom}, with
$U$'s being replaced by ${\cal U}$'s), and has the form
\be
\label{betabb}
B \sim {\cal U}_{\bar{b}}^{a} \bar{\cal U}_{a}^{\bar{b}} =
{\cal N}_{\bar {c} \bar{b}}^{a} \bar{\cal N}_{c a}^{\bar{b}} e_\mu^c {\bar e}_\nu^{\bar{c}}
 \partial X^{\mu}\bar{\partial} X^{\nu}
\ee
which is another form of the result \rf{betab}, and we have introduced in \rf{betabb} the tangent-space
components ${\cal N}$ of the Nijenhuis tensor.

To understand the last equality in \rf{betabb},
consider connection in the complexified tangent
bundle with torsion
\be
\label{decab}
de^{a}= (a_\mu)^{a}_{c} dX^\mu\wedge e^c + (b_\mu)^{a}_{\bar{c}} dX^\mu\wedge
{\bar e}^{\bar{c}}
\ee
The Nijenhuis tensor comes from the $b$-part of \rf{decab} to be
defined as
\be\label{dn}
[  {\bar e}_{\bar{b}} , {\bar e}_{\bar{c}} ] = - {\cal N}_{\bar{b}
\bar{c}}^{a} e_a +{\sf f}_{\bar{b} \bar{c}}^{\bar{a}}{\bar e}_{\bar{a}}
\ee
where $e_a$ and $\bar{e}_{\bar{b}}$ are considered as vector fields
and equation (\ref{dn}) can be considered as integrability condition of the
system of equations on holomorphic functions
\be\label{comp}
{\bar e}_{\bar{b}} f = {\bar e}_{\bar{b}}^{\mu}\frac{\partial}{\partial X^\mu} f = 0,\ \ \
{\bar b}=1,\ldots,D/2
\ee
On the equations of motion $e^a_\mu{\bar\d}X^\mu=0$ for the theory \rf{foa} one can write
\be
{\cal U}_{\bar{b}}^{a}=(de)_{\mu\nu}^{a} e_{\bar{b}}^{\mu}
\bar{\partial} X^{\nu} = (de)_{\mu\nu}^{a} e_{\bar{b}}^{\mu}{\bar e}^\nu_{\bar c}
{\bar e}_\lambda^{\bar c}
\bar{\partial} X^{\lambda}
\ee
This gives rise immediately to \rf{betabb} since
\be
{\bar e}_{\bar a}^\mu\d_\mu{\bar e}^\nu_{\bar b} = - {\bar e}_{\bar a}^\mu\left(
{\bar e}^\nu_{\bar c}\d_\mu{\bar e}^{\bar c}_\lambda{\bar e}^\lambda_{\bar b}
+ e^\nu_c\d_\mu e^c_\lambda{\bar e}^\lambda_{\bar b}
\right)
\ee
The first term in the r.h.s. is inessential for the Nijenhuis tensor in \rf{dn},
while the second gives
\be
\label{BN}
(de)^{a}_{\mu\nu}e^\mu_{\bar b}{\bar e}^\nu_{\bar c} = \d_{[\mu}e^a_{\nu]}e^\mu_{\bar b}
{\bar e}^\nu_{\bar c}=-{\bar e}^{\nu}_{[\bar{b}}\d_{\nu}{\bar e}^{\mu}_{\bar{c}]}e^{a}_{\mu}
= {\cal N}_{\bar{c} \bar{b}}^{a}
\ee
which completes the proof of \rf{betabb}. The bets function \rf{betabb} in the first-order
theory \rf{foa} is again expressed in terms of the Nijenhuis tensor and therefore vanishes on the
Kodaira-Spencer equations.

Let us finally present explicit relations between the vielbein language and original first-order
theory. Solving \rf{inviel} for the inverse vielbein at the point, where
$e^{a}_{j} = \delta^a_j$ and $e^{a}_{\bar{j}} = \mu^a_{\bar{j}}$, one gets
 \be
 e^{i}_{a} = M^{i}_a,\,\,\,\,\,\,
 e^{i}_{\bar{b}} = - M^{i}_{k}\mu^{k}_{\bar{b}} = - M^{\bar{k}}_{\bar{b}}\mu^{i}_{\bar{k}}
 \ee
and the rest can be obtained by complex conjugation. Hence,
 \be
 {\bar e}_{\bar{b}} = \bar{M}^{\bar{k}}_{\bar{b}}\mathcal{D}_{\bar{k}}
 \ee
where the long derivatives \rf{longder} give rise to the Nijenhuis tensor
by their commutator $[\mathcal{D}_{\bar i},\mathcal{D}_{\bar j}] = -N^k_{\bar i\bar j}\d_k$.
Therefore
\be
\label{NN}
 [{\bar e}_{\bar{b}} , {\bar e}_{\bar{c}}] =
 [\bar{M}^{\bar{k}}_{\bar{b}}\mathcal{D}_{\bar{k}},\bar{M}^{\bar{l}}_{\bar{c}}\mathcal{D}_{\bar{l}}] = -
 \bar{M}^{\bar{k}}_{\bar{b}}\bar{M}^{\bar{l}}_{\bar{c}}N^{a}_{\bar{k}\bar{l}}\d_a +
 \bar{M}^{\bar{k}}_{[\bar{b}}\mathcal{D}_{\bar{k}}\bar{M}^{\bar{l}}_{\bar{c}]}\mathcal{D}_{\bar{l}} =
 \\
 =  - \bar{M}^{\bar{k}}_{\bar{b}}\bar{M}^{\bar{l}}_{\bar{c}}N^{a}_{\bar{k}\bar{l}}e_a +
 ( \bar{M}^{\bar{k}}_{[\bar{b}}\mathcal{D}_{\bar{k}}\bar{M}^{\bar{l}}_{\bar{c}]}-
 \bar{M}^{\bar{k}}_{\bar{b}}\bar{M}^{\bar{s}}_{\bar{c}}N^{a}_{\bar{k}\bar{s}}M^{\bar{l}}_{\bar{d}}\bar{\mu}^{\bar{d}}_{a})\mathcal{D}_{\bar{l}}
   \ee
since $\d_i= M^{k}_i\mathcal{D}_k + M^{\bar{k}}_{\bar{l}}\bar{\mu}^{\bar{l}}_{i}\mathcal{D}_{\bar{k}}$.
Comparing \rf{NN} with \rf{dn}, one finds
 \be
 \bar{M}^{\bar{k}}_{\bar{b}}\bar{M}^{\bar{l}}_{\bar{c}}N^{a}_{\bar{k}\bar{l}} =
 \mathcal{N}^{a}_{\bar{b}\bar{c}}
 \ee
The symmetries in vielbein formalism are more transparent.
Consider the perturbed action:
\be
L[e,g,b] =
p_ae^a_{\mu}\bar{\d}X^{\mu}+c.c. + g^{a\bar{a}}p_ap_{\bar{a}} + b_{\mu\nu}\d X^{\mu}\bar{\d} X^{\nu}
\ee
Under the general co-ordinate transformation
\be
X^{\mu} \rightarrow X^{\mu} - v^{\mu}(X)
\ee
which is an equivalent of \rf{gencoord}, this action transforms into:
\be
\delta L[e,g,b] = \left[(\delta p_a)e^a_{\mu}+
p_a(\delta e^a_{\mu}-v^{\nu}\d_{\nu}e^a_{\mu}-e^a_{\nu}\d_{\mu}v^{\nu})\right]\bar{\d}X^{\mu}+
c.c. + \delta (g^{a\bar{a}}p_ap_{\bar{a}}) -
\\
- v^{\mu}\d_{\mu}g^{a\bar{a}}p_ap_{\bar{a}}
+ \left[\delta b_{\mu\nu} -
v^{\rho}\d_{\rho}b_{\mu\nu} - b_{\rho\nu}\d_{\mu} v^{\rho}-b_{\mu\rho}\d_{\nu} v^{\rho}\right]\d X^{\mu}\bar{\d} X^{\nu}
\ee
Thus, it is invariant, in particular, under the following transformations
\be
\delta p_a = 0,\,\,\, \delta g^{a\bar{a}} = v^{\mu}\d_{\mu}g^{a\bar{a}}
\\
\delta b_{\mu\nu} =
v^{\rho}\d_{\rho}b_{\mu\nu} + b_{\rho\nu}\d_{\mu} v^{\rho}+b_{\mu\rho}\d_{\nu} v^{\rho}
\\
\delta e^{a}_{\mu} = v^{\nu}\d_{\nu}e^a_{\mu}+e^a_{\nu}\d_{\mu}v^{\nu}
\ee
and the ${\cal N}$-components of the Nijenhuis tensor simply transform as
 \be
 \delta \mathcal{N}^{a}_{\bar{b}\bar{c}} = v^{\rho}\d_{\rho} \mathcal{N}^{a}_{\bar{b}\bar{c}}
 \ee

\section{Discussion}

We have considered in this paper the first-order theory, which hypothetically
corresponds to a string theory
in singular backgrounds, when expanding around the bare action \rf{act}.
This theory depends only on the (local) choice of target-space complex structure and we have
studied this dependence by perturbing the bare action by the Beltrami vertex operator
\rf{Fimu}.

The background effective action, which arises after dividing world-sheet fields into the
slow and fast variables and the one-loop integration over the latter contains the beta-function
of the operator \rf{vb}, vanishing on the Kodaira-Spencer equations \rf{KoSpe} for the target-space
Beltrami differentials \rf{vmu}. This result comes after an almost trivial computation
of the Gaussian integral, which
however clarifies several delicate issues, and requires certain field redefinition
in the space of fast variables, corresponding
to change of complex polarization in target-space for a nontrivial background.
The computed beta-function is proportional to {\em squared} Kodaira-Spencer equations
\rf{betab},
where the conjugated components of the Nijenhuis tensor \rf{KoSpe}
are contracted by certain matrices, depending on
the components of Zamolodchikov metric is the space of primary operators or background fields
for the first-order string theory. Reformulating this result in the form \rf{betabb}, one finds
that this extra metric in the space of target-space fields can be rewritten in the form of
inverse vielbein fields \rf{inviel}.
We have also discussed briefly the consistency of this result
with the exact symmetries of the perturbed theory.

This result can be also reproduced from direct computation of the correlation functions in
conformal theory \rf{act}. The ``co-ordinate'' beta-functions arise in such
approach as particular integrals
over the moduli spaces of punctured world sheets of the first-order string theory, partially
this approach has been discussed in \cite{Cardy}.
The detailed analysis of this approach will be published in the second part of this
paper \cite{GM2}, and here finally we will only sketch the idea.

To study the co-ordinate approach to the beta-functions consider, for example, the perturbed
one-point correlation function of the ``probe operator'' \rf{vg}
\be
\label{gfi}
\langle O_g(x)\rangle_t = \langle O_g(x)\exp (t\int_\Sigma\Phi)\rangle =
\sum_{n\geq 0}{t^n\over n!}\int_\Sigma d^2z_1\ldots\int_\Sigma d^2z_n
\langle  O_g(x)\Phi(z_1)\ldots\Phi(z_n)\rangle
\ee
where averaging in the r.h.s. is understood in the sense of path integral with the free
action \rf{act}. The calculation of the r.h.s. of \rf{gfi} includes the integration of
the multipoint correlators
\be
\label{corint}
\langle O_g(x)\Phi(z_1)\ldots\Phi(z_n)\rangle
\ee
over the regularized domain $\Sigma^{\otimes n}$, e.g.
\be
\label{dom}
|z_i-x|>\epsilon,\ \ \ \ |z_i-z_j|>\epsilon
\\
\forall i,j=1,\ldots,n
\ee
In order to get a hint of what should we expect from such computation, consider the first
nontrivial order, namely the correlator \rf{corint} for $n=2$
\be
\label{gfifi}
\langle O_g(x)\Phi(y)\Phi(z)\rangle =
\langle O_g(x)O_\mu(y) O_{\bar\mu}(z)\rangle +
\langle O_g(x)O_{\bar\mu}(y) O_\mu(z)\rangle
\ee
The direct computation of the free field correlator in the r.h.s. of \rf{gfifi}
gives rise to the result
\be\label{C2}
\frac{1}{2}\langle O_g(x) \int_{\Sigma}\Phi(y)\int_{\Sigma}\Phi(z)\rangle \approx \frac{g^{i\overline{j}}B^{(2)}_{i\overline{j}}}{|x|^4}\log\frac{R^2}{\e^2}
\int\limits_{|y|<R} \frac{d^2y}{\pi}
\ee
The coefficient at the logarithmic singularity \rf{C2}
is proportional to the function
\be
\label{ks0}
g^{i{\bar j}}B^{(2)}_{i{\bar j}} =
g^{k{\bar k}}\d_{[i}{\bar\mu}^{\bar j}_{k]}\d_{[{\bar k}}\mu^i_{{\bar j}]}
\ee
Strictly speaking, instead of
$g^{k{\bar k}}\d_{[i}{\bar\mu}^{\bar j}_{k]}\d_{[{\bar k}}\mu^i_{{\bar j}]}$ one should write
a target-space integral
\be
\label{tarint}
\int d^D X^{(0)}g^{k{\bar k}}(X^{(0)})\d_{[i}{\bar\mu}^{\bar j}_{k]}(X^{(0)})
\d_{[{\bar k}}\mu^i_{{\bar j}]}(X^{(0)})
\ee
over the zero modes $X^{(0)}$ of the $X$-fields,
and consider the compact target, or the target-space fields being
coefficients functions of the operators
\rf{vg} and \rf{vmu} vanishing at the space-time ``infinity'', what allows integration
by parts in \rf{tarint}; together with the transversality constraints
\rf{transv} this bring us to \rf{gfifi}.
The computation of the 3-point function \rf{gfifi} is almost equivalent to the
calculation of the operator product expansion of two operators $\Phi$
\be
  \label{OOPE}
  \Phi(z)\Phi(0) = {2{\bar\mu}_i^{\bar j}\mu^i_{\bar j}\over |z|^4} +
  {\d_k\left({\bar\mu}_i^{\bar j}\mu^i_{\bar j}\right)\d X^k\over z{\bar z}^2} +
  {\d_{\bar k}\left({\bar\mu}_i^{\bar j}\mu^i_{\bar j}\right){\bar\d} X^{\bar k}\over z^2{\bar z}}
  + \frac{2 {\tilde B}^{(2)}_{i\bar{j}}\d X^i\bar{\d} X^{\bar{j}}}{|z|^2} + \ldots
\\
{\tilde B}^{(2)}_{i\bar{j}} = B^{(2)}_{i\bar{j}} +
\half\left(\d_i{\sf w}_{\bar{j}} + \d_{\bar{j}}{\sf w}_i\right)
\\
{\sf w}_{\bar{j}}= \d_{[\bar{j}}\mu^k_{\bar{k}]}\bar{\mu}^{\bar{k}}_{k},\,\,\,\,
    {\sf w}_i = \d_{[i}\bar{\mu}^{\bar{k}}_{k]}\mu^k_{\bar{k}}
   \ee
with its further projection onto the operator $O_g$, being in this sense
equivalent to the computation of the quadratic contribution into the beta-function
of the operator $O_b$ (see e.g. \cite{Pol,Zam}).

We see therefore, that computation of the singularity of the 3-point function \rf{gfifi}
reproduces the quadratic in $\mu$-fields piece of the beta-function \rf{betab}.
It is quite instructive to discuss therefore, how the next terms $B^{(3)}_{i{\bar j}}$ arise from the
four-point contribution. This is already not very trivial computation, requiring special care,
when considering the integrals over the world-sheet moduli space \cite{GM2}.

It is important, that nonlinear equations for the
background fields arise in this approach from the correlation functions with {\em different} number of
operators (the so called polyvertex structures, already discussed in this context
in \cite{LMZ,Zeitl}),
and contain integrals over the moduli spaces of punctured world sheets with different numbers
of punctures. This is close to already discussed in similar context structures of the
string field theory (see e.g. \cite{Zwi,Son}).
Generally we believe that such approach should lead to solvable nonlinear equations for
the ``co-ordinate'' beta-functions and we are going to return to these issues elsewhere.

\bigskip\noindent
{\bf Acknowledgements}

\noindent
We are grateful to G.~Arutyunov, M.~Kontsevich, N.~Nekrasov, A.~Rosly, A.~Smilga
and A.~Tseytlin for important discussions. OG also thanks V.~Shadura and N.~Iorgov
for permanent encouragement and support. AM thanks the IHES in Bures-sur-Yvette,
the MPIM in Bonn and the Galileo Galilei Institute in Florence, where essential
parts of this work have been done.

The work of OG and ASL was supported by the grant for support of Scientific Schools
LSS-3036.2008.2, the work of OG was supported by French-Ukrainian program ``Dnipro'', the
project  M17-2009, the joint project PICS of CNRS and Nat. Acad. Sci. of Ukraine and the project
F28.2/083 of FRSF of Ukraine. The work of ASL was also supported by RFBR grant 07-01-00526.
The work of AM was supported by the Russian Federal Nuclear Energy Agency,
the RFBR grant 08-01-00667,
the grant for support of Scientific Schools LSS-1615.2008.2, the
INTAS grant 05-1000008-7865, the project ANR-05-BLAN-0029-01, the
NWO-RFBR program 047.017.2004.015, the Russian-Italian RFBR program 09-01-92440-CE,
the Russian-Ukrainian RFBR program 09-02-90493, the Russian-French RFBR-CNRS program
09-02-93105 and by the
Dynasty foundation.


\end{document}

\end{document}